\documentclass{article}%
\usepackage{amsmath}
\usepackage{amssymb}
\usepackage{amsfonts}
\usepackage{graphicx}
\usepackage{hyperref}
\usepackage{mathpazo}%
\providecommand{\U}[1]{\protect\rule{.1in}{.1in}}
\begin{document}

\begin{center}
{\Large Branched Hamiltonians and Supersymmetry}\bigskip

{\large T L Curtright}$^{\S }${\large \ and C K Zachos}$^{\natural}$\bigskip

$^{\S }$Department of Physics, University of Miami, Coral Gables, FL
33124-8046, USA\\[0pt]\textsl{curtright@miami.edu}\bigskip

$^{\natural}$High Energy Physics Division, Argonne National Laboratory,
Argonne, IL 60439-4815, USA \\[0pt]\textsl{zachos@anl.gov}\bigskip
\end{center}

\centerline{\bf Abstract} \phantom{bunga} \textit{Some examples of branched
Hamiltonians are explored both classically and in the context of quantum
mechanics, as recently advocated by Shapere and Wilczek. \ These are in fact
cases of switchback potentials, albeit in momentum space, as previously
analyzed for quasi-Hamiltonian chaotic dynamical systems in a classical
setting, and as encountered in analogous renormalization group flows for
quantum theories which exhibit RG cycles.} \ \textit{A basic two-worlds model,
with a pair of Hamiltonian branches related by supersymmetry, is considered in
detail.}

\section{Introduction}

Multi-valued Hamiltonians have appeared in at least two contexts. \ Most
recently, they have resulted from Legendre transforming Lagrangians whose
velocity dependence is not convex \cite{SW1,SW2}, which invariably leads to a
Riemann surface phase-space structure, with multiply-branched Hamiltonians,
and to interesting topological issues \cite{SWX,Wil}. \ Previously, they have
arisen in the continuous interpolation of discrete time dynamical systems,
particularly those systems that exhibited chaotic behavior, where they could
be incorporated in a canonical \textquotedblleft
quasi-Hamiltonian\textquotedblright\ formalism \cite{CZ,CZchaos,CV,C}. \ 

Moreover, by analogy with quasi-Hamiltonian systems, renormalization group
flows that exhibit cycles have also been shown to be governed by multi-valued
$\beta$ functions \cite{CZrg,CJZ}.

We consider here several simple Lagrangian models that lead to double-valued
Hamiltonian systems, to illuminate \textquotedblleft
two-worlds\ theory.\textquotedblright\ \ We begin with an example where the
velocity dependence of $L$ is given by a gaussian. \ This example illustrates
many generic features of branched Hamiltonians, in addition to its more
specific peculiarities. \ In particular, as a quantum system the gaussian
model is not amenable to solution in closed form, so we turn to a different
class of models where analytic results can be obtained. One of the models in
this class is tailored so as to have a pair of Hamiltonians that comprise a
supersymmetric quantum mechanical system \cite{Wit}. \ This facilitates
obtaining analytic results as well as numerical study of this special model.

\section{A gaussian model with momentum switchbacks}

For an interesting example, consider a non-convex $v$-dependent gaussian
Lagrangian:%
\begin{align}
L\left(  x,v\right)   &  =C\left(  1-\exp\left(  -\frac{1}{2C}~mv^{2}\right)
\right)  -V\left(  x\right)  \ ,\label{Gaussian L(v)}\\
p\left(  v\right)   &  =\frac{\partial L}{\partial v}=mv\exp\left(  -\frac
{1}{2C}~mv^{2}\right)  \ . \label{Gaussian p(v)}%
\end{align}
Most of what can be said about this model can be stated at the classical
level. $\ L$ is a union of three convex functions defined on the three $v$
intervals $\left(  -\infty,-\sqrt{C/m}\right]  $, $\left[  -\sqrt{C/m}%
,\sqrt{C/m}\right]  $, and $\left[  \sqrt{C/m},\infty\right)  $.
\ The\ width\ parameter $C$ sets the energy scale. \ When plotted versus $v$,
the kinetic energy of the model has the classic shape of a \emph{fedora }hat
profile.%
\begin{center}
\includegraphics[
height=2.9888in,
width=4.4858in
]%
{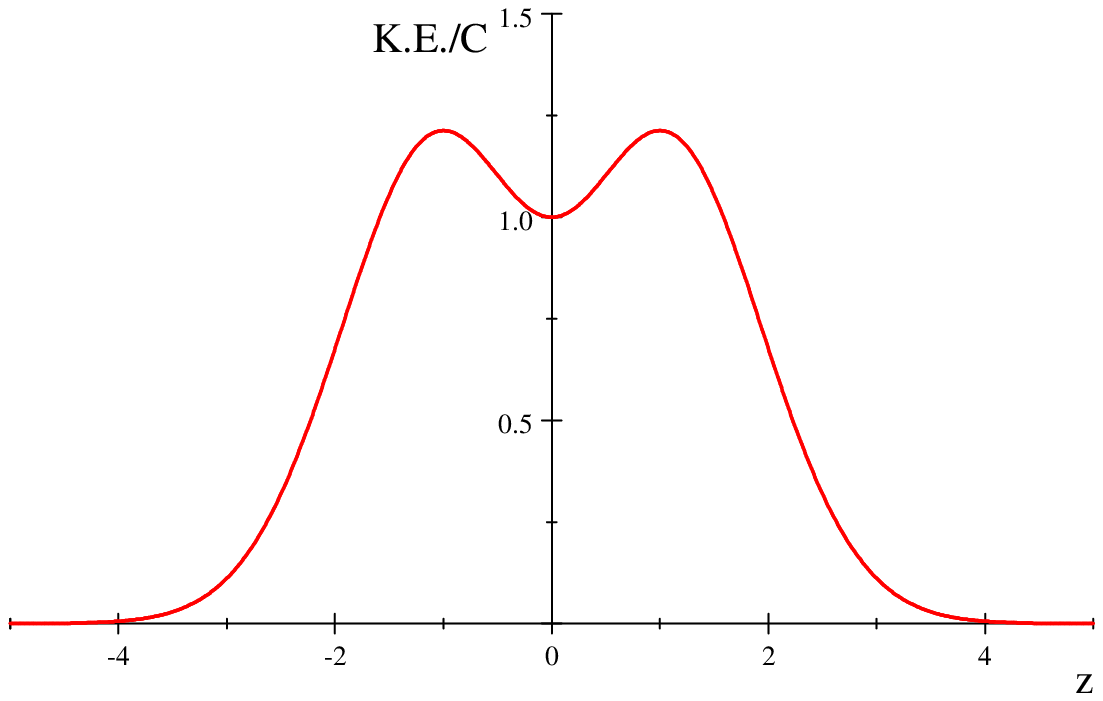}%
\\
Kinetic energy, $\left(  L+V\right)  /C=\left(  1+\frac{mv^{2}}{C}\right)
e^{-\frac{1}{2}\frac{mv^{2}}{C}}$, versus $z\equiv v\sqrt{m/C}$, for the
gaussian model.
\end{center}

For this model, $v$ and $p$ always have the same sign, and clearly
$-\infty\leq v\leq+\infty$. \ However, due to the gaussian suppression in $v$,
the momentum $p$ is confined to a \emph{finite} interval, as given by the
maximum and minimum of (\ref{Gaussian p(v)}), namely, $\left.  p\left(
v\right)  \right\vert _{v=\pm\sqrt{C/m}}=\pm\sqrt{mC/e}$. \ 

Moreover, there are two values for $H$ at every value of $p\in\left(
-\sqrt{mC/e},\sqrt{mC/e}\right)  $. \ To see this double-valued $H$, we invert
(\ref{Gaussian p(v)}) to obtain%
\begin{equation}
v\left(  p\right)  =\pm\sqrt{-\frac{C}{m}\operatorname{LambertW}\left(
-\frac{p^{2}}{mC}\right)  }\ ,
\end{equation}
where both real branches of the negatively-valued Lambert function, for
negative argument, are allowed. \ Thus the Hamiltonian, $H\left(  x,p\right)
=p~v\left(  p\right)  -L\left(  x,v\left(  p\right)  \right)  $, as a function
of position and momentum, is%
\begin{align}
H\left(  x,p\right)   &  =\sqrt{\frac{Cp^{2}}{m}}\left(  \pm\sqrt
{-\operatorname{LambertW}\left(  -\frac{p^{2}}{mC}\right)  }\pm\frac{1}%
{\sqrt{-\operatorname{LambertW}\left(  -\frac{p^{2}}{mC}\right)  }}\right)
-C+V\left(  x\right) \nonumber\\
&  =V\left(  x\right)  +\frac{1}{2m}~p^{2}+\frac{1}{8Cm^{2}}~p^{4}+\frac
{5}{48m^{3}C^{2}}~p^{6}+O\left(  p^{8}\right)  \ , \label{Gaussian H(p)}%
\end{align}
where the low momentum expansion is valid near $p=0$ for the upper, principal
branch of the Lambert function.

Now, since there are \emph{two} real branches for both the square-root
function and the Lambert function, we might expect four values for $H$ at any
given momentum. \ However, the square-root and LambertW branches are always
correlated, as is evident upon considering the $\left(  p\left(  v\right)
,H\left(  x,p\left(  v\right)  \right)  \right)  $ curve in parametric form on
the $\left(  p,H\right)  $ plane, using $v$ as the parameter, so that the
gaussian model's Hamiltonian is only double-valued for all $p\in\left(
-\sqrt{mC/e},\sqrt{mC/e}\right)  $. \ This is shown in the following Figure
for $V\left(  x\right)  =0$. \ Note the Hamiltonian curve closes, as a
function of $p$, with three cusps.%
\begin{center}
\includegraphics[
height=2.9915in,
width=4.4905in
]%
{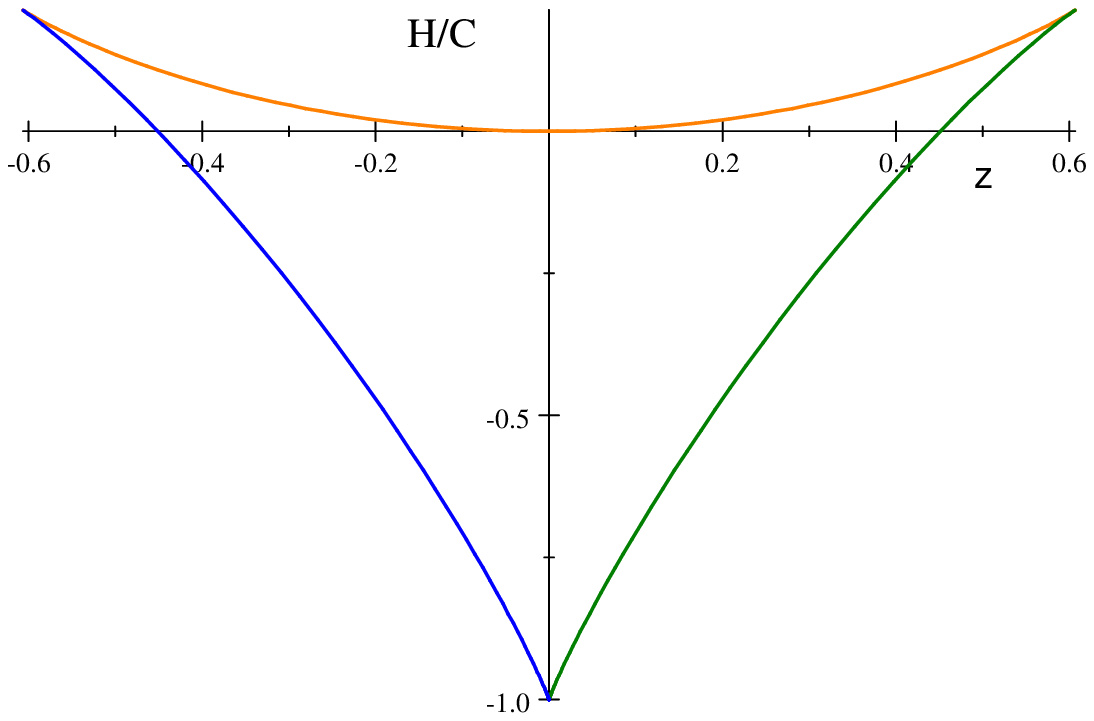}%
\\
The real branches of $H/C$ versus $z\equiv p/\sqrt{mC}\in\left[  -1/\sqrt
{e},1/\sqrt{e}\right]  \approx\left[  -0.61,0.61\right]  $.
\end{center}
While only double-valued, $H$ is clearly the union of \emph{three} convex
functions, defined on three overlapping momentum intervals: $\ H_{-}$, $H_{0}%
$, and $H_{+}$ for $p\in\left[  -\sqrt{mC/e},0\right]  $, $\left[
-\sqrt{mC/e},\sqrt{mC/e}\right]  $, and $\left[  0,\sqrt{mC/e}\right]  $, as
displayed in the Figure in blue, orange, and green, respectively.\ \ 

Classical trajectories for this model, given a specific choice for $V\left(
x\right)  $, evince the switchback potential phenomena discussed at length in
\cite{CZchaos}, only here in momentum rather than position space. \ 

For instance, selecting the harmonic potential, $V\left(  x\right)
=C+\frac{1}{2}m\omega^{2}x^{2}$, it is straightforward to plot trajectory
curves in terms of either $\left(  x,v\right)  $ or $\left(  x,p\right)  $.
\ Some explicit $\left(  x,p\right)  $ phase space trajectories for various
fixed energies are plotted below (for $\frac{1}{2}m\omega^{2}=1=C$). \ More
information is available
\href{http://server.physics.miami.edu/~curtright/MiamiFedoraTrajectories.pdf}{online}%
, where trajectories are also shown on the $\left(  x,v\right)  $
configuration surface (a cylinder, actually). \ 

When moving on a trajectory governed by one branch of $H$, a classical
particle will encounter one of the Hamiltonian cusps in finite time, in
general, and then bounce (switch) to be governed by another branch of $H$.
\ Because of this switching, trajectories may intersect and cross in the
Figure. \ This cannot happen for a system governed by a single-valued
Hamiltonian, as is well-known, but it is allowed when different Hamiltonian
branches are governing the motion for the different curves that cross.\ \ A
system governed by a multi-valued Hamiltonian usually does exhibit this novel
feature. \ We have called such trajectories \textquotedblleft
quasi-Hamiltonian\textquotedblright\ flows\ in our earlier work \cite{CZchaos}.

The unified 3-fold structure of $H$ brings to mind some previous theories
exhibiting triality \cite{Shankar}, along with supersymmetry. \ However, to
our knowledge the gaussian model above shows no compelling signs of
supersymmetry. \ Still, it would be quite interesting to find a simple,
three-Hamiltonian, single-particle quantum system, based on a single unifying
Lagrangian, that could be partitioned into pairs of supersymmetric
Hamiltonians, with state-linking operators of a type familiar from
supersymmetric quantum mechanics. \ 

In the following Sections, we will analyze a different model with a
double-valued Hamiltonian that \emph{does} exhibit supersymmetry. \ But first
some preliminaries. \ The gaussian model does not readily admit analytic,
closed form results when quantized, even with so simple a potential as
$V\left(  x\right)  =C+\frac{1}{2}m\omega^{2}x^{2}$, so we turn to a class of
models where exact quantum results can be more easily obtained.%
\begin{center}
\includegraphics[
height=4.3275in,
width=6.4904in
]%
{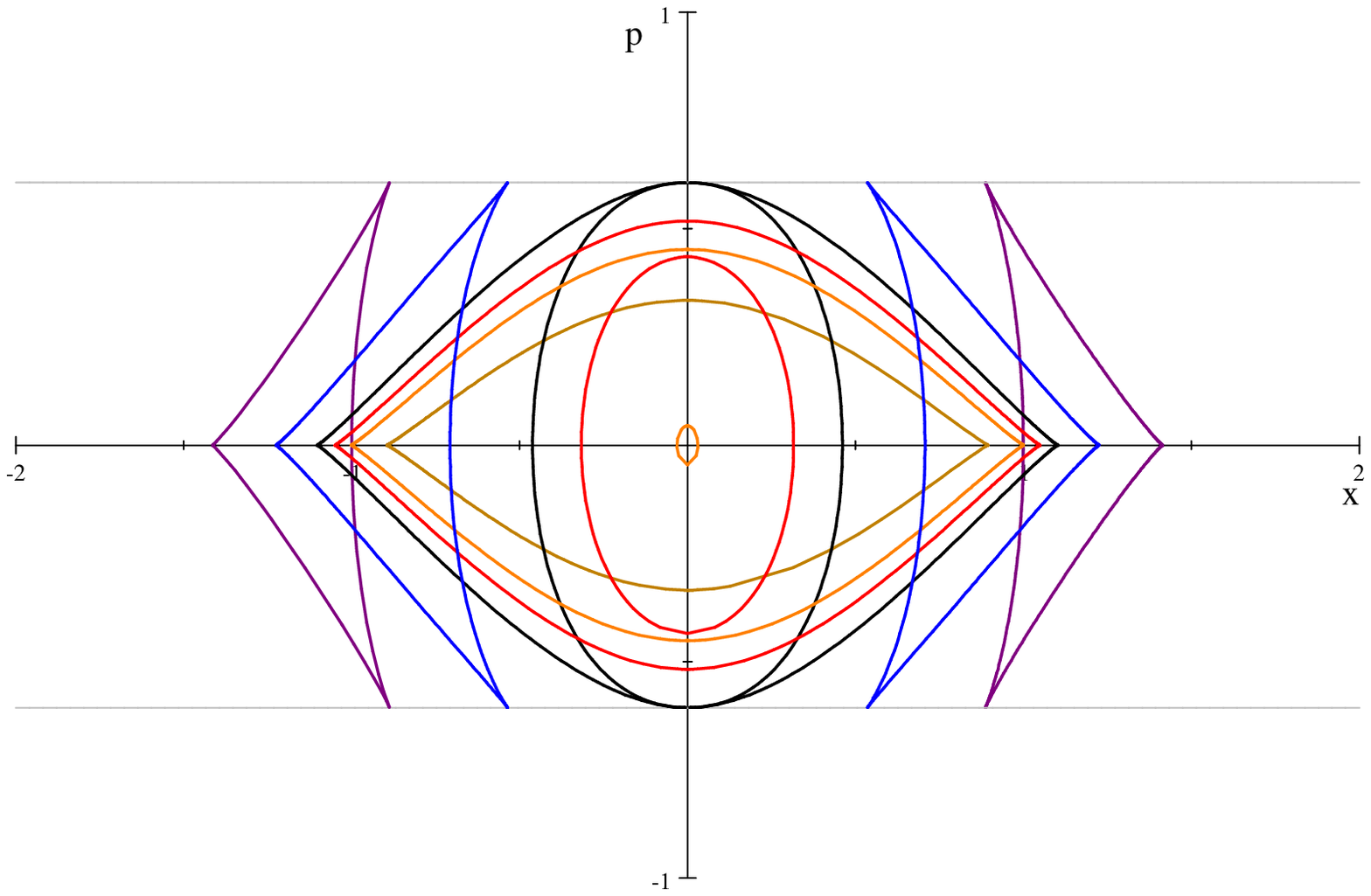}%
\\
Gaussian model phase space trajectories for $E=2/\sqrt{e}\approx1.21$,
$E=3/2$, \& $E=2$ are shown in black, blue, \& purple, respectively, and also
for $E=0.800,\ 1.001,\ $\& $1.100$, in sienna, orange, \& red, respectively.
\ The black curves constitute a separatrix. \ The outer, black, cusped curve
is approached from within by trajectories whose $E$ is increased from $0$ to
$2/\sqrt{e}$, whle the inner, black oval is approached not only from without
by the cusped, triangular trajectories, as $E$ is decreased to $2/\sqrt{e}$,
but also from within by bounded, closed oval orbits, as $E$ is increased from
$1$ to $2/\sqrt{e}$.
\end{center}

\section{A class of double-valued Hamiltonians}

For positive integer $k$, consider\footnote{This class of models could be
generalized to $L=C\left(  v-c\right)  ^{n/m}-V\left(  x\right)  $ for any
fixed $c$ \& $C$, and for any odd integer $n$ \& odd integer $m$. \ There
seems to be no real gain or simplification achieved by doing so, except
perhaps for the choice $c=0$.}%
\begin{equation}
L=C\left(  v-1\right)  ^{\frac{2k-1}{2k+1}}-V\left(  x\right)  \text{ \ \ with
\ \ }C\equiv\frac{2k+1}{2k-1}\left(  \frac{1}{4}\right)  ^{\frac{2}{2k+1}%
}>0\ . \label{L}%
\end{equation}
For real $v$ we take the $1/\left(  2k+1\right)  $ roots to be real, such that
$\left(  v-1\right)  ^{\frac{1}{2k+1}}\gtrless0$ for $v\gtrless1$. \ By doing
this we are in fact taking the real parts of two different branches of the
analytic $1/\left(  2k+1\right)  $\textsl{\ }roots as a function of complex
$v$. \ We do this solely to have a real, single-valued Lagrangian function for
\emph{all} real $v$.

So far as we can tell, there is no particularly compelling reason not to draw
on more than one branch of an analytic function of $v$ so long as only one
branch is encountered at any given real $v$, or at least that would seem to be
true for classical dynamics. \ We will discuss the consequences this choice
for $L$ has for the quantum dynamics in the following, especially for the case
$k=1$.%
\begin{center}
\includegraphics[
height=2.8401in,
width=4.256in
]%
{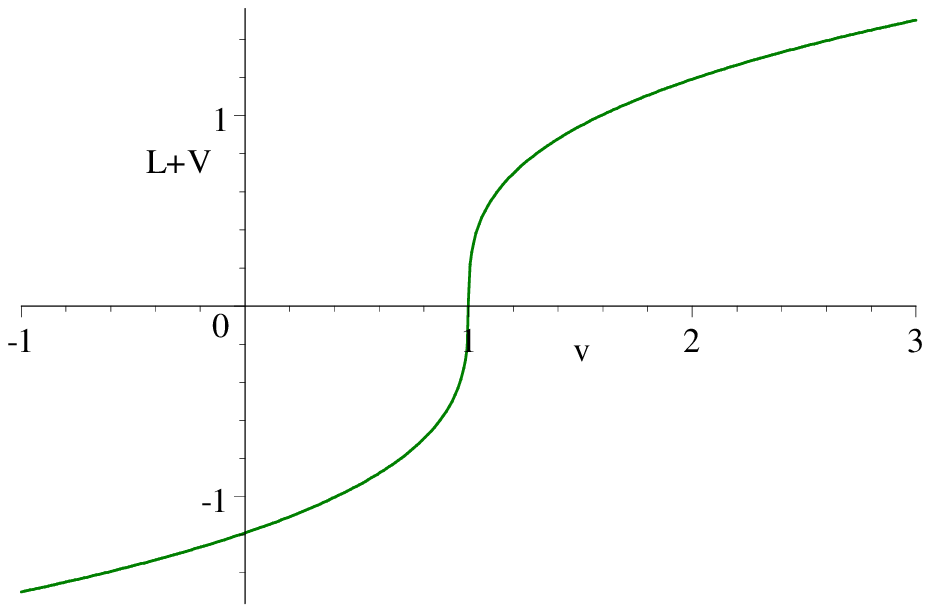}%
\\
The $k=1$ case, $L+V=C\left(  v-1\right)  ^{\frac{1}{3}}$.
\end{center}

For $v$ near zero, we then have $L\approx C\left(  -1+v~\frac{2k-1}%
{2k+1}+v^{2}~\frac{2k-1}{\left(  2k+1\right)  ^{2}}+O\left(  v^{3}\right)
\right)  $. \ Of these terms, the first is innocuous, the second would give a
boundary contribution to the action and therefore not effect the equations of
motion, and the third is the usual $v^{2}$ kinetic structure:
\begin{align}
A  &  =\int_{t_{1}}^{t_{2}}Ldt\nonumber\\
&  \approx C\left(  t_{2}-t_{1}+\frac{2k-1}{2k+1}\left(  x\left(
t_{2}\right)  -x\left(  t_{1}\right)  \right)  +\frac{2k-1}{\left(
2k+1\right)  ^{2}}\int_{t_{1}}^{t_{2}}v^{2}dt+\int_{t_{1}}^{t_{2}}O\left(
v^{3}\right)  dt\right) \nonumber\\
&  -\int_{t_{1}}^{t_{2}}V\left(  x\right)  dt\ .
\end{align}
So this action would yield the usual Newtonian classical equations of motion
for small $v$. \ On the other hand, for large velocities, the $v$ dependence
is more elaborate, leading (for finite, positive integer $k$) to a non-convex
function of velocity, whose curvature $\partial^{2}L/\partial v^{2}$ flips
sign at just one point, namely, $v=1$. \ 

Thus, the function $L$ may be thought of a single \emph{pair} of convex
functions judiciously pieced together. \ The non-convexity of $L$\ has the
effect of making the kinetic energy, and hence the Hamiltonian, a
\emph{double-valued} function of $p$. \ For any positive integer $k$, we find
\emph{two} branches for $H$,%
\begin{equation}
H_{\pm}=p\pm\frac{1}{4k-2}\left(  \frac{1}{\sqrt{p}}\right)  ^{2k-1}+V\left(
x\right)  \ . \label{H+-}%
\end{equation}
This follows from
\begin{equation}
p=\partial L/\partial v=\left(  \frac{1}{4}\right)  ^{\frac{2}{2k+1}}~\frac
{1}{\left[  \left(  v-1\right)  ^{2}\right]  ^{1/2k+1}}\ , \label{p(v)}%
\end{equation}
whose inverse $v\left(  p\right)  $ is double-valued,
\begin{equation}
v_{\pm}\left(  p\right)  \equiv1\mp\frac{1}{4}\left(  \frac{1}{\sqrt{p}%
}\right)  ^{2k+1}\ . \label{v(p)}%
\end{equation}
The pair of Hamiltonians in (\ref{H+-}) are then obtained by taking the
\emph{Legendre transform} with respect to each of the two $v$ branches,%
\begin{equation}
H_{\pm}\left(  x,p\right)  =pv_{\pm}\left(  p\right)  -L\ ,
\end{equation}
where we have used $L\left(  x,p\right)  =\mp\frac{1}{4}\frac{2k+1}%
{2k-1}\left(  \frac{1}{\sqrt{p}}\right)  ^{2k-1}-V\left(  x\right)  $ on the
$v_{\pm}\left(  p\right)  $ branches. \ 

For $k=1$, the two kinetic energy branches have the shape shown in the Figure
below. \ Note that, \emph{classically}, $p$ must be non-negative for this
model to avoid imaginary $v\left(  p\right)  $. \ That is to say, the slope
$\partial L/\partial v$ is always positive.%
\begin{center}
\includegraphics[
height=2.8401in,
width=4.256in
]%
{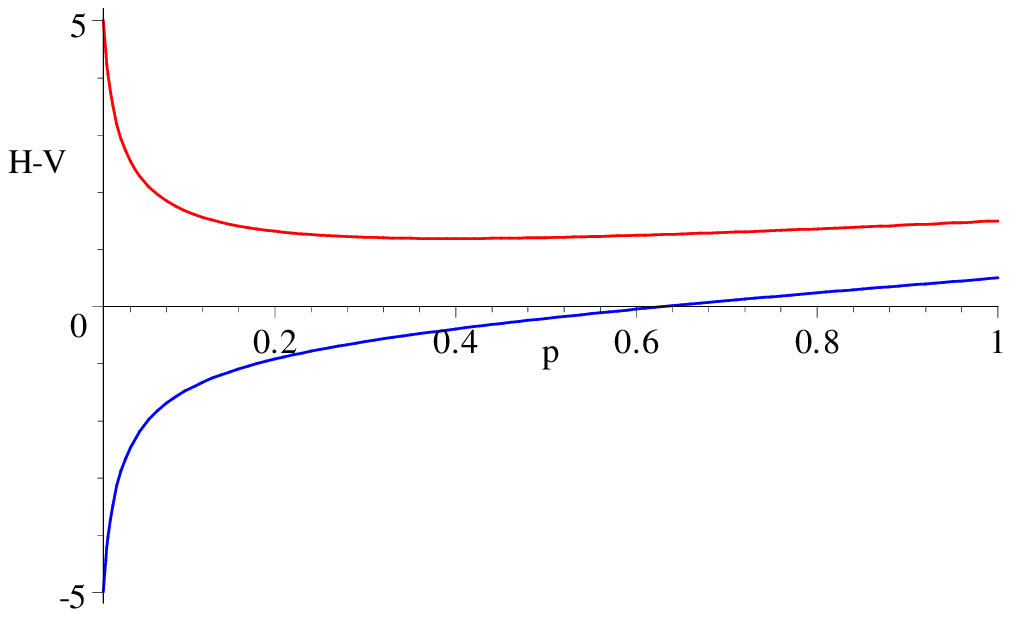}%
\\
$\left.  H_{\pm}-V\left(  x\right)  \right\vert _{k=1}=p\pm\frac{1}{2\sqrt{p}%
}$ in red/blue. \ There is a cusp at $p=\infty$ where both $H_{\pm
}\protect\underset{p\rightarrow\infty}{\sim}p$.
\end{center}

Following the suggestions of Shapere and Wilczek \cite{SW2}, we define the
associated \emph{quantum} theory with $p\geq0$ as a restriction, with various
boundary conditions imposed on the wave functions, $\psi\left(  p\right)  $,
at $p=0$, such that there is no probability flow to negative $p$.

\section{A supersymmetric model}

We purposefully plotted the $k=1$ case of (\ref{L}), and related quantities,
in the above Figures. \ The $k=1$ case is special when the potential $V\left(
x\right)  $\ is harmonic: \ It is a \emph{supersymmetric} quantum mechanical
system when viewed in momentum space. \ In that case, $C=3/4^{2/3}\approx1.19$
and%
\begin{gather}
L=C\left(  v-1\right)  ^{1/3}-V\left(  x\right)
\ \ \ \underset{\text{Legendre}}{\Longleftrightarrow}\ \ \ H_{\pm}=p\pm
\frac{1}{2\sqrt{p}}+V\left(  x\right)  \ ,\\
V\left(  x\right)  =x^{2}\underset{\text{QM in }p\text{ space}}{\longmapsto
}-\frac{d^{2}}{dp^{2}}\ .
\end{gather}

\subsection{Quantum features}

The momentum space pair of QM Hamiltonian operators for this case is therefore
expressible in the standard form for a supersymmetric pair,%
\begin{equation}
H_{\pm}=-\frac{d^{2}}{dp^{2}}+w_{0}^{2}\left(  p\right)  \pm w_{0}^{\prime
}\left(  p\right)  =\left(  \frac{d}{dp}\pm w_{0}\left(  p\right)  \right)
\left(  -\frac{d}{dp}\pm w_{0}\left(  p\right)  \right)  \ , \label{superH+-}%
\end{equation}
where $w_{0}\left(  p\right)  =\sqrt{p}$. \ This has the interesting feature
that the true --- square-integrable --- ground state of the system is
non-vanishing for only one of the branches, namely, $H_{-}$. \ 

As an algebraic system, for $p\geq0$, the two Hamiltonians are related in a
familiar fashion by
\begin{gather}
H_{-}=a^{\dag}a\ ,\ \ \ H_{+}=aa^{\dag}=H_{-}+\left[  a,a^{\dag}\right]
=H_{-}+1/\sqrt{p}\ ,\\
a\equiv\frac{d}{dp}+\sqrt{p}\ ,\ \ \ a^{\dag}\equiv-\frac{d}{dp}+\sqrt
{p}\ ,\ \ \ \left[  a,a^{\dag}\right]  =\frac{1}{\sqrt{p}}\ .
\end{gather}
Both energy spectra are non-negative given either Dirichlet or Neumann
boundary conditions at $p=0$.\footnote{There is a subtlety here. \ Strictly,
$a^{\dag}$ is \emph{not} the adjoint of $a$. \ For states subject to Neumann
conditions, $\left.  \psi\right\vert _{p=0}\neq0=\left.  \psi^{\prime
}\right\vert _{p=0}$, there are nonvanishing boundary contributions at $p=0$:
$\int_{0}^{\infty}\left(  \psi_{2}\left(  a^{\dag}\psi_{1}^{\ast}\right)
-\psi_{1}^{\ast}\left(  a\psi_{2}\right)  \right)  dp=\left.  \psi_{1}^{\ast
}\psi_{2}\right\vert _{p=0}$. \ Nevertheless, it is still true for $\psi$
satisfying \emph{either} Neumann \emph{or} Dirichlet conditions that
$\left\langle H_{-}\right\rangle =\int_{0}^{\infty}\left\vert a\psi\right\vert
^{2}dp\geq0$ and $\left\langle H_{+}\right\rangle =\int_{0}^{\infty}\left\vert
a^{\dag}\psi\right\vert ^{2}dp\geq0$, because for such states, $\left.
\psi^{\ast}\left(  a\psi\right)  \right\vert _{p=0}=0=\left.  \psi^{\ast
}\left(  a^{\dag}\psi\right)  \right\vert _{p=0}$.} \ 

The zero-energy ground state of $H_{-}$ is given by%
\begin{equation}
a\psi_{0}\left(  p\right)  =0\ ,\ \ \ \psi_{0}\left(  p\right)  =N_{0}%
\exp\left(  -\frac{2}{3}p^{3/2}\right)  \ ,\ \ \ N_{0}=\frac{6^{1/6}}%
{\sqrt{\Gamma\left(  \frac{2}{3}\right)  }}\approx1.16\ .
\label{SuperGroundState}%
\end{equation}
This obeys the boundary condition $\psi_{0}^{\prime}\left(  0\right)
=0\neq\psi_{0}\left(  0\right)  $, and is normalized such that $\int%
_{0}^{\infty}\left\vert \psi_{0}\left(  p\right)  \right\vert ^{2}dp=1$. \ 

On the other hand, the zero-energy state for $H_{+}$, namely, $\phi\left(
x\right)  =\exp\left(  +\frac{2}{3}p^{3/2}\right)  $, is not admissible,
because it has infinite norm.

The higher energy states are degenerate, with $H_{\pm}\psi^{\left(
\pm\right)  }=E\psi^{\left(  \pm\right)  }$ eigenstates for $E>0$ mutually
related by
\begin{equation}
\psi_{E}^{\left(  +\right)  }=\frac{1}{\sqrt{E}}~a\psi_{E}^{\left(  -\right)
}\ ,\ \ \ \psi_{E}^{\left(  -\right)  }=\frac{1}{\sqrt{E}}~a^{\dag}\psi
_{E}^{\left(  +\right)  }\ ,
\end{equation}
so as to have equal norms. \ In particular the first excited state for $H_{-}
$ is degenerate with the lowest energy state for $H_{+}$, with $E_{1}=1.89379
$, as determined by numerical analysis. \ \ 

All this conforms with well-known expectations for general supersymmetric QM.
\ Due to the restriction $p\geq0$, there is perhaps an interesting wrinkle
here, albeit previously encountered for the supersymmetric simple harmonic
oscillator (but normally expressed in terms of $\psi\left(  x\right)  $):
\ The degenerate $H_{\pm}$ eigenfunctions obey different boundary conditions
at $p=0$. \ If one is Dirichlet, the other is Neumann. \ This follows from the
mutual relations between $\psi_{E}^{\left(  \pm\right)  }$ and the fact that
$a$ and $a^{\dag}$ reduce to $\pm d/dp$ when acting on nonsingular functions
at $p=0$. \ For example, the first $H_{-}$ excited state and its degenerate
$H_{+}$ partner eigenstate satisfy $\left.  \psi_{E_{1}}^{\left(  -\right)
}\right\vert _{p=0}=0=\left.  d\psi_{E_{1}}^{\left(  +\right)  }/dp\right\vert
_{p=0}$, while for the next excited states, $\left.  d\psi_{E_{2}}^{\left(
-\right)  }/dp\right\vert _{p=0}=0=\left.  \psi_{E_{2}}^{\left(  +\right)
}\right\vert _{p=0}$, etc. \ 

Flipping the boundary conditions actually has a practical benefit due to the
$1/\sqrt{p}$ singularity in both $H_{\pm}$: \ It is more straightforward to
perform an accurate numerical computation of the energy eigenvalue using the
boundary condition $\psi_{E}\left(  0\right)  =0\neq\psi_{E}^{\prime}\left(
0\right)  $ than it is using the condition $\psi_{E}\left(  0\right)
\neq0=\psi_{E}^{\prime}\left(  0\right)  $. \ The degeneracy of the
eigenfunctions permits one to always choose the $\psi_{E}\left(  0\right)  =0$
condition, along with the corresponding $H_{+}$ or $H_{-}$.

These higher energy states may be thought of as a single nontrivial state
defined on a unified covering space --- a double covering of the half-line $%
\mathbb{R}
_{+}$ by $%
\mathbb{R}
$ --- obtained by unfolding the two Hamiltonian branches to obtain a single
$H$ \cite{SW2} globally defined on $%
\mathbb{R}
$. \ However, as is clear from the preceding discussion, the true ground state
of the system is $\psi_{0}\left(  p\right)  $ $\cup\ 0$ on the unfolded space.
\ The latter, somewhat unusual feature is possible because the two
Hamiltonians on the half-lines join together in a cusp at $p=\infty$, where
$\psi_{0}$ and all its derivatives vanish. \ So too vanish all the higher
$\psi_{E}^{\left(  \pm\right)  }$ and all their derivatives at $p=\infty$. \ 

For this reason, it would be excusable not to have thought of the degenerate
eigenstates on the half-line as two branches of a single function. \ However,
the unified two-worlds picture provided by joining them together on a covering
real line, with Neumann and Dirichlet boundary conditions at opposite ends, is
a more compelling point of view, in our opinion. \ Perhaps more importantly,
this omniscient view of the two-worlds system becomes natural when the common
Lagrangian underpinning both $H_{\pm}$ is considered.

\paragraph*{Self-adjointness of $H$ and probability flow.}

For arbitrary superpositions of momentum space wave functions, $\psi=\sum
_{n}c_{n}\psi_{n}$, with each of the $\psi_{n}$ obeying either Dirichlet or
Neumann boundary conditions, the Hamiltonians are real but not self-adjoint.
\ While the behavior at $p=\infty$ is sufficiently benign for all normalizable
linear combinations of energy eigenfunctions, the behavior at $p=0$\ could be
a problem since%
\begin{equation}
\int_{0}^{\infty}\left(  \left(  H_{\pm}\chi^{\ast}\right)  \psi-\chi^{\ast
}\left(  H_{\pm}\psi\right)  \right)  dp=\left.  \psi\frac{d}{dp}\chi^{\ast
}\right\vert _{p=0}-\left.  \chi^{\ast}\frac{d}{dp}\psi\right\vert _{p=0},
\end{equation}
and this does not necessarily vanish. \ To avoid this and ensure
self-adjointness of $H$, a superselection rule may be imposed \emph{\cite{SSR}%
}: \ The Hilbert space may be partitioned into Dirichlet and Neumann
solutions, $\mathcal{H}=\mathcal{H}_{D}\oplus\mathcal{H}_{N}$, while allowing
no mixing of the two. \ Thus, superpositions of only Dirichlet or only Neumann
wave functions are permitted, but linear combinations of both are not. \ 

While restrictive, this rule nevertheless retains the novel double-valued
Hamiltonian feature of the model.\ \ Both branches of $H_{\pm}$ are operative
on each of $\mathcal{H}_{D}$\ and $\mathcal{H}_{N}$. \ Linear combinations of
progressively higher energy levels that alternate between $H_{-}$ and $H_{+}$
eigenstates, i.e. $\psi=\sum_{n}b_{n}\psi_{E_{n}}^{\left(  -\right)  }%
+c_{n}\psi_{E_{n+1}}^{\left(  +\right)  }$, maintain the self-adjointness of
$H$, while requiring both $H_{\pm}$ for their time evolution.

The same superselection rule guarantees conservation of probability at $p=0$.
\ Wave packets in either $\mathcal{H}_{D}$ or $\mathcal{H}_{N}$ will not
transport probability to negative $p$.

\subsection{Classical features}

It is also instructive to survey essential features of the classical
trajectories for the model. \ The Euler-Lagrange equations are%
\begin{equation}
\frac{dv}{dt}=\frac{9}{C}~x~\sqrt[3]{\left(  v-1\right)  ^{5}}\ . \label{EofM}%
\end{equation}
where $9/C=3\times4^{2/3}\approx7.\,56$. \ So we immediately see there are
\emph{special solutions}: \ $v=1$ for any initial $x$ results in $v\left(
t\right)  =1$ for all $t$. \ Therefore, for any $x\left(  0\right)  $,
\begin{equation}
x\left(  t\right)  =x\left(  0\right)  +t\ .
\end{equation}

More generally, if $v>1$ at any time, then it will remain so for all $t$, the
force does not restore, and $x$ will grow with $t$, faster than exponential in
fact. \ In this case the time it takes for $x$ to go to $\infty$ is finite.
\ But if $v<1$ at any time, it will remain so for all $t$, the force is
restoring, and the solution oscillates, albeit nonlinearly. \ 

Classically, energy conservation along a given configuration space trajectory
may be expressed as constant $E$ where%
\begin{equation}
E\left(  x,v\right)  =x^{2}+\frac{C}{3}\frac{3-2v}{\sqrt[3]{\left(
v-1\right)  ^{2}}}\ . \label{K}%
\end{equation}
Note that this $E$ is \emph{single}-valued as a function of $v$, even though
$H_{\pm}\left(  x,v\right)  =vp_{\pm}\left(  v\right)  -L\left(  x,v\right)  $
is \emph{double}-valued as a function of $v$. \ 

How can this be? \ It is possible just because the two branches of $H_{\pm
}\left(  x,v\right)  $ appear on \emph{opposite} sides of $v=1$, and not for
the same value of $v$. \ That is to say, it all comes back to our choice for
the cube roots on the real line. \ By taking $L$ real for both $v>1$ and for
$v<1$, we have in fact used two different branches of the analytic cube root
function defined for complex $z$. \ However, with our construction, we
encounter only one branch of this analytic function, and hence one value of
$L$, at any given real value of $v$. \ The story is different for the two
Hamiltonians, $H_{\pm}\left(  x,p\right)  $, where we encounter both branches
for every $p>0$.

Upon detailed inspection of constant $E\left(  x,v\right)  $ curves on the
$\left(  x,v\right)  $ plane, one finds that, for $E<C$, there are only open
trajectories with $v>1$, and all these escape to $x=\infty$ in finite times,
in which case only $H_{-}\left(  x,p\right)  $ is operative; while for $E\geq
C $, on the $\left(  x,v\right)  $ plane there are not only open, unbounded
trajectories, for all $v>1$, as governed again by $H_{-}\left(  x,p\right)  $,
but also closed, bounded trajectories, for all $v<1$, as governed by
$H_{+}\left(  x,p\right)  $. \ 

Hence for $E\geq C$ classical trajectories exist in which both $H_{+}$ and
$H_{-}$ are operative. \ This should be compared with the existence of
admissible wave functions for both $H_{\pm}$ with identical energy eigenvalues
$E>C$.

In fact, the $E<C$ classical situation provides intuition that is in accord
with the features of the quantum ground state. \ For $E<C$, there are
\emph{no} classical trajectories (whether open and unbounded, or closed and
bounded) in the $v<1$ region governed by $H_{+}\left(  x,p\right)  $. \ For
$E<C$, rather, there are only unbounded classical trajectories in the $v>1$
region governed by $H_{-}\left(  x,p\right)  $. \ Hence, for $E<C$, a path
integral of $\exp\left(  iA/\hbar\right)  $ would encounter no stationary
points if restricted to trajectories in the $v<1$ region. \ Moreover, there is
an infinite, impenetrable $E$ barrier separating classical solutions with
$v<1$ from those with $v>1$, as is evident in the Figure below. \ This would
suggest there are no admissible wave functions for $E<C$ with support in the
region $v<1$. \ Or, in terms of $p$, for energy less than $C$, there would be
no admissible $\psi\left(  p\right)  $ energy eigenstates governed by $H_{+}$.
\ This heuristic argument is in agreement with the quantum features of the
model.%
\begin{center}
\includegraphics[
height=2.84in,
width=4.2557in
]%
{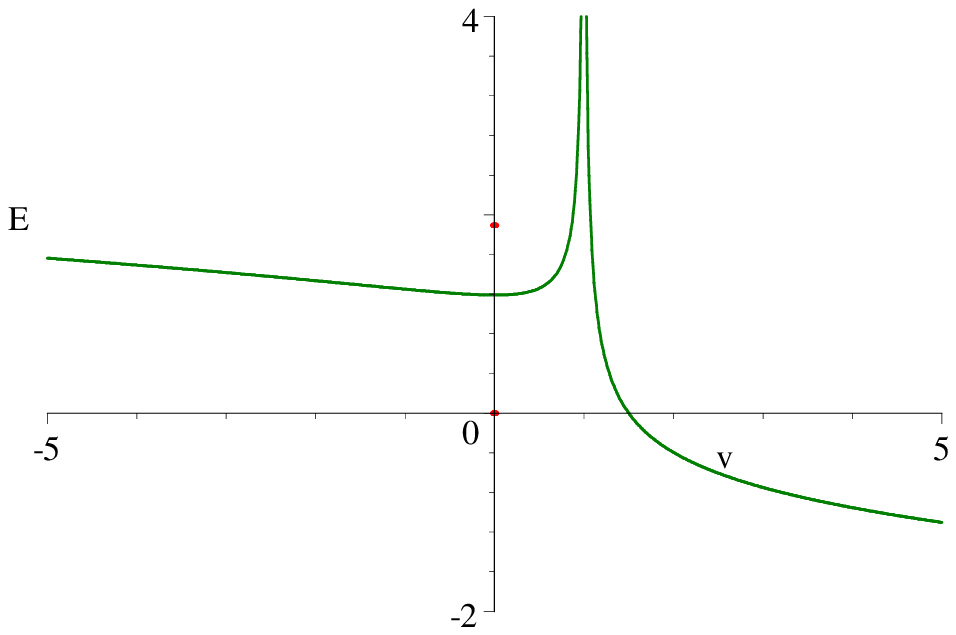}%
\\
$E\left(  x,v\right)  $ as given by (\ref{K}) is clearly minimized on the
$\left(  x,v\right)  $ plane along the line $x=0$, and less obviously for
$v<1$ at $v=0$. \ This is evident in a graph of $\ \left.  E\left(
x,v\right)  \right\vert _{x=0}$ versus $v$. \ The minimum for $v<1$ is at the
$\left(  x,v\right)  $ origin, where $E\left(  0,0\right)  =C=\frac{3}%
{4^{2/3}}\approx1.19$. \ The red dots on the $E$ axis are the lowest two
energy eigenvalues for the quantized model, namely, $E_{0}=0$ and
$E_{1}=1.89379$.
\end{center}

Perhaps these classical features underpinning the quantized model become
clearer upon considering trajectories as constant energy curves in $\left(
x,p\right)  $ phase space. \ Several representative examples are shown in the
next Figure (more details are available
\href{http://server.physics.miami.edu/~curtright/SuperDuperTrajectories.pdf}{online}%
). \ Two energies shown in the Figure allow both open, unbounded trajectories,
and closed, bounded orbits, namely, for $E=1.2$ and $E=1.4$. \ As noted in the
previous Figure, there is a critical energy, $E=C=\frac{3}{4^{2/3}}%
\approx1.\,19$, below which bounded orbits do not occur.

When bounded orbits do exist, their turning points are given by $x=\pm
\sqrt{E-C}$, corresponding to $v=0$ in (\ref{K}). \ However, at these turning
points the momentum does \emph{not} vanish, being given instead by
$p=C/3=\frac{1}{4^{2/3}}\approx0.397$, as indicated by the horizontal light
gray line in the Figure.%
\begin{center}
\includegraphics[
height=3.6841in,
width=5.5253in
]%
{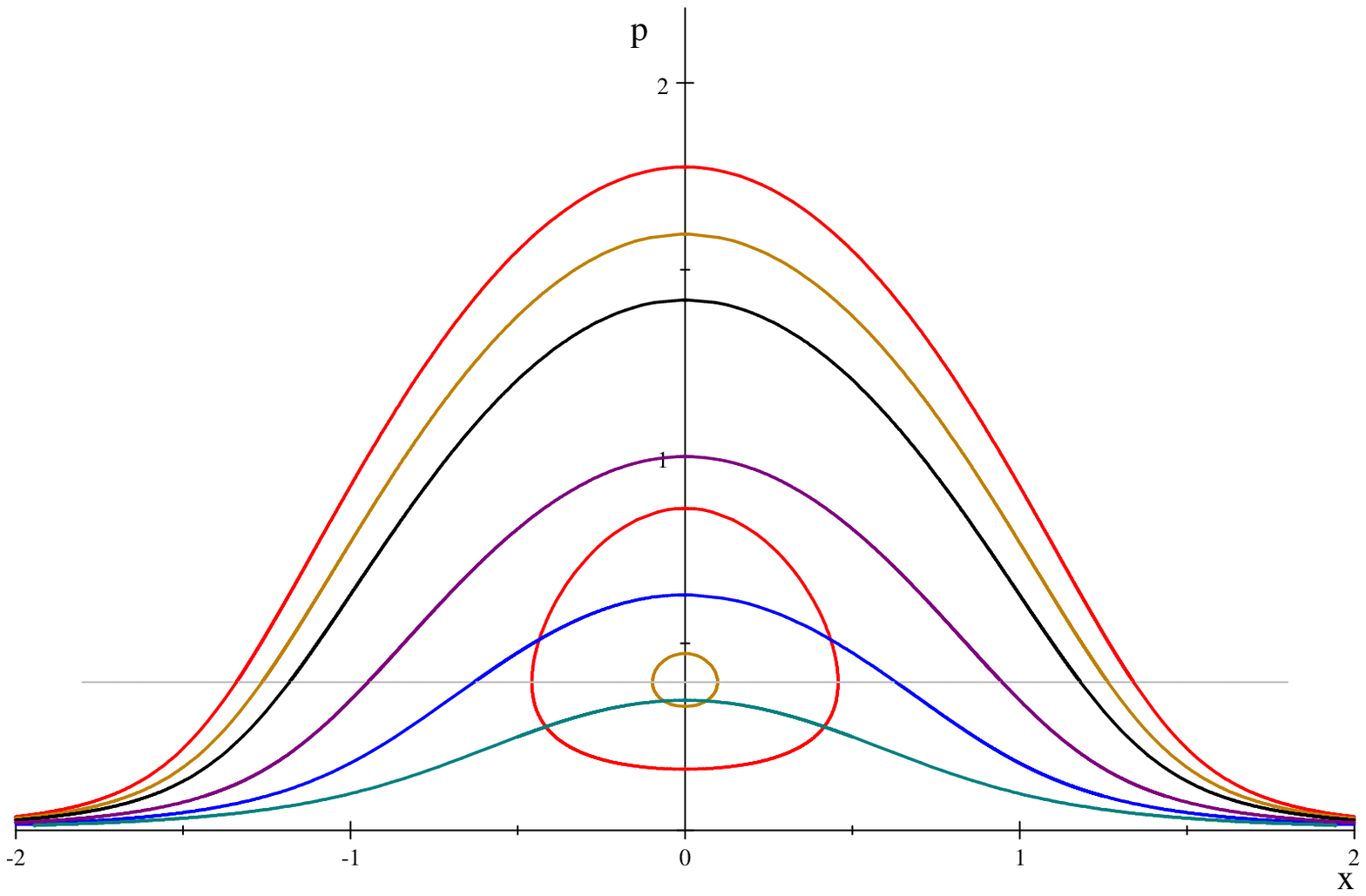}%
\\
Supersymmetric model phase space trajectories are shown for various energies:
\ $E=-0.5$ in blue-green, $E=0$ in blue, $E=0.5$ in purple, $E=1$ in black,
$E=1.2$ in sienna, and $E=1.4$ in red.
\end{center}

It is important to note in the Figure the counter-intuitive feature that $p>0$
even when $v\leq0$. \ Moreover, the phase space curves also exhibit
quasi-Hamiltonian flow \cite{CZchaos}, as mentioned above for the gaussian
model: \ Trajectories can cross each other on this $\left(  x,p\right)  $
phase space plot. \ This is allowed when different Hamiltonian branches are
governing the motion for the different curves that cross. \ That is to say,
just like $H$, the trajectories are actually on two different branches of a
phase space Riemann surface.

\section{Deforming the supersymmetric Hamiltonians}

Here we implement a deformation procedure \cite{M,R} to construct a family of
related but modified Hamiltonians through the use of general solutions to the
Riccati equation as obtained from particular solutions.

In addition to the square-integrable zero-energy solution of $H_{-}\psi=0$, as
given by
\begin{equation}
\left(  \frac{d}{dp}+\sqrt{p}\right)  \psi\left(  p\right)  =0\text{ \ \ i.e.
\ \ }\psi\left(  p\right)  =\exp\left(  -2p^{3/2}/3\right)  \ ,
\end{equation}
the factorized Hamiltonian method may be used to construct another
square-integrable solution, for a\emph{\ modified} Hamiltonian, from the
\emph{non}-square-integrable zero-energy solution of $H_{+}\phi=0$, as given
by%
\begin{equation}
\left(  \frac{d}{dp}-\sqrt{p}\right)  \phi\left(  p\right)  =0\text{ \ \ i.e.
\ \ }\phi\left(  p\right)  =\exp\left(  2p^{3/2}/3\right)  \ .
\end{equation}

The construction involves the general solution of the Riccati equation
$V_{-}=w^{2}-w^{\prime}$, as obtained from the particular one used above,
$w_{0}\left(  p\right)  =\sqrt{p}$. \ This general solution involves the
non-square-integrable $\phi$, and a single constant of integration, $\kappa$.
\ Thus%
\begin{align}
w_{\kappa}\left(  p\right)   &  =w_{0}\left(  p\right)  -\frac{d}{dp}%
\ln\left(  1+\kappa\int_{0}^{p}e^{2\int_{0}^{s}w_{0}\left(  u\right)
du}ds\right)  \nonumber\\
&  =w_{0}\left(  p\right)  -\frac{\kappa e^{2\int_{0}^{p}w_{0}\left(
u\right)  du}}{1+\kappa\int_{0}^{p}e^{2\int_{0}^{s}w_{0}\left(  u\right)
du}ds}\ .
\end{align}
For the case at hand, this comes down to%
\begin{align}
w_{\kappa}\left(  p\right)   &  =\sqrt{p}-\frac{\kappa e^{4p^{3/2}/3}%
}{1+\kappa g\left(  p\right)  }\label{w(p)}\\
&  =\sqrt{p}-\kappa e^{4p^{3/2}/3}+\kappa^{2}e^{4p^{3/2}/3}g\left(  p\right)
+O\left(  \kappa^{3}\right)  \ ,
\end{align}%
\begin{equation}
g\left(  p\right)  \equiv\int_{0}^{p}e^{4s^{3/2}/3}ds=pe^{\frac{4}{3}%
p^{\frac{3}{2}}}\left.  _{1}F_{1}\right.  \left(  1;5/3;-\frac{4}{3}%
p^{\frac{3}{2}}\right)  \ .\label{g(p)}%
\end{equation}
So the subleading terms in this $\kappa$-deformation involve a confluent
hypergeometric function (incomplete gamma). \ Note that $w_{0}\left(
p\right)  =\left.  w_{\kappa}\left(  p\right)  \right\vert _{\kappa=0}$.
\ Also note that $w_{\kappa}^{2}-w_{\kappa}^{\prime}=p-\frac{1}{2\sqrt{p}}$
for any $\kappa$.

Now, a zero-energy eigenfunction of $H_{-}$ constructed from the general
$w_{\kappa}$ is \emph{not} square-integrable (except in the case $\kappa=0$).
\ However, a new square-integrable solution for a \emph{modified}
Hamiltonian\emph{\ }$H_{+}\left(  \kappa\right)  $ can be constructed. \ 

For any $\kappa>0$ this solution is%
\begin{equation}
\phi_{0}\left(  p,\kappa\right)  =\frac{\kappa e^{\int_{0}^{p}w_{0}\left(
u\right)  du}}{1+\kappa\int_{0}^{p}e^{2\int_{0}^{s}w_{0}\left(  u\right)
du}ds}\ ,
\end{equation}
where it is significant that the exponent in the numerator is one half that in
$w_{\kappa}$. \ For the present case this is%
\begin{equation}
\phi_{0}\left(  p,\kappa\right)  =\frac{\kappa e^{\frac{2}{3}p^{\frac{3}{2}}}%
}{1+\kappa g\left(  p\right)  }\ .
\end{equation}
Note that this solution disappears in the undeformed limit, $\left.  \phi
_{0}\left(  p,\kappa\right)  \right\vert _{\kappa=0}=0$. \ Also note the
square-integrability on the half-line, $0\leq p\leq\infty$: \ This holds for
the $\phi_{0}$ wave functions, for all $\kappa>0$. \ In the next Figure, we
plot some representative $\phi_{0}\left(  p,\kappa\right)  $ for selected
$\kappa$.

\paragraph{\textbf{Boundary conditions.}}

For general $\kappa$, $\phi_{0}$ satisfies neither Neumann nor Dirichlet, but
rather Robin boundary conditions\footnote{For example, see
http://en.wikipedia.org/wiki/Robin\_boundary\_condition} depending on $\kappa
$, namely, $\ $%
\begin{equation}
\kappa\phi_{0}\left(  0,\kappa\right)  +d\phi_{0}\left(  0,\kappa\right)
/dp=0\ .
\end{equation}
This follows from \ $\phi_{0}\left(  0,\kappa\right)  =\kappa\ ,\ \ \ d\phi
_{0}\left(  0,\kappa\right)  /dp=-\kappa^{2}$.%
\begin{center}
\includegraphics[
height=2.8401in,
width=4.256in
]%
{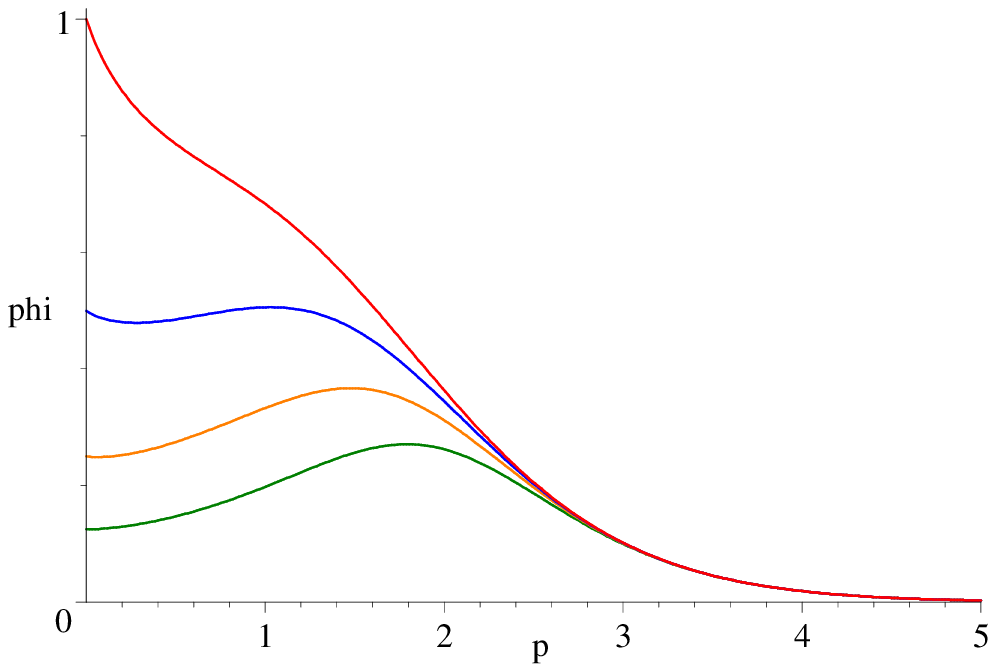}%
\\
$\phi_{0}\left(  p,\kappa\right)  $ for $\kappa=1,\ 1/2,\ 1/4,\ $\& $1/8$, in
red, blue, orange, \& green, respectively. \ Note that $\phi_{0}\left(
p,0\right)  =0$.
\end{center}

Of course, it is better to write the derivative of $\phi_{0}$ at all values of
$p\geq0$ as a linear null equation:%
\begin{equation}
\left(  \frac{d}{dp}-w_{\kappa}\left(  p\right)  \right)  \phi_{0}\left(
p,\kappa\right)  =0\ .
\end{equation}
In this form it is clear that $\phi_{0}\left(  p,\kappa\right)  $ is a
square-integrable zero-energy solution of a $\kappa$-dependent class of
Hamiltonians involving the general $w_{\kappa}\left(  p\right)  $:%
\begin{align}
H_{+}\left(  \kappa\right)   &  =-\left(  \frac{d}{dp}+w_{\kappa}\left(
p\right)  \right)  \left(  \frac{d}{dp}-w_{\kappa}\left(  p\right)  \right) \\
&  =-\frac{d^{2}}{dp^{2}}+p+\frac{1}{2\sqrt{p}}-\frac{4\kappa\sqrt
{p}e^{4p^{3/2}/3}}{1+\kappa g\left(  p\right)  }+\frac{2\kappa^{2}%
e^{8p^{3/2}/3}}{\left(  1+\kappa g\left(  p\right)  \right)  ^{2}}\ .\nonumber
\end{align}
Note, then, that $\left.  H_{+}\left(  \kappa\right)  \right\vert _{\kappa
=0}=H_{+}$, the initial undeformed Hamiltonian, as given in equation
(\ref{superH+-}).

By way of comparison, $H_{-}\left(  \kappa\right)  =-\left(  \frac{d}%
{dp}-w_{\kappa}\left(  p\right)  \right)  \left(  \frac{d}{dp}+w_{\kappa
}\left(  p\right)  \right)  $ does not participate in this deformation, as it
is actually \emph{independent} of $\kappa$, and identical to the previous
$H_{-} $ in equation (\ref{superH+-}). \ As mentioned earlier, in this case
the $\kappa$-dependent zero energy eigenfunctions of $H_{-}$, as given by
$\exp\left(  -\int_{0}^{p}w_{\kappa}\left(  s\right)  ds\right)  $, are
\emph{not} square-integrable except for $\kappa=0$. \ 

That is to say, the true ground state of $H_{-}$ is indeed unique, and
proportional to $\exp\left(  -2p^{3/2}/3\right)  $. \ The normalization of the
true ground state is finite and given by $\int_{0}^{\infty}\exp\left(
-4p^{3/2}/3\right)  dp=\frac{1}{6^{1/3}}~\Gamma\left(  \frac{2}{3}\right)
\approx0.745$. \ On the other hand, it is informative to check that
$\exp\left(  -\int_{0}^{p}w_{\kappa}\left(  s\right)  ds\right)  $ is
\emph{not} square integrable for $\kappa>0$, where%
\begin{equation}
w_{\kappa}\left(  p\right)  =\sqrt{p}-\frac{\kappa e^{4p^{3/2}/3}}{1+\kappa
g\left(  p\right)  }\ .
\end{equation}
To see this, it is sufficient just to plot $w_{\kappa}$ for a few values of
$\kappa$ and infer the general result. \ 

For any $\kappa>0$, it is evident from the Figure below that $w_{\kappa}$
becomes negative and grows in magnitude for large enough $p$, asymptoting
towards a common $\kappa$-independent function in the limit of large $p$.
\ Thus we have $\int_{0}^{p}w_{\kappa}\left(  u\right)  du<0$ for $p$
sufficiently large, and hence $\int_{0}^{\infty}\exp\left(  -2\int_{0}%
^{p}w_{\kappa}\left(  s\right)  ds\right)  dp$ will diverge.%
\begin{center}
\includegraphics[
height=2.84in,
width=4.2557in
]%
{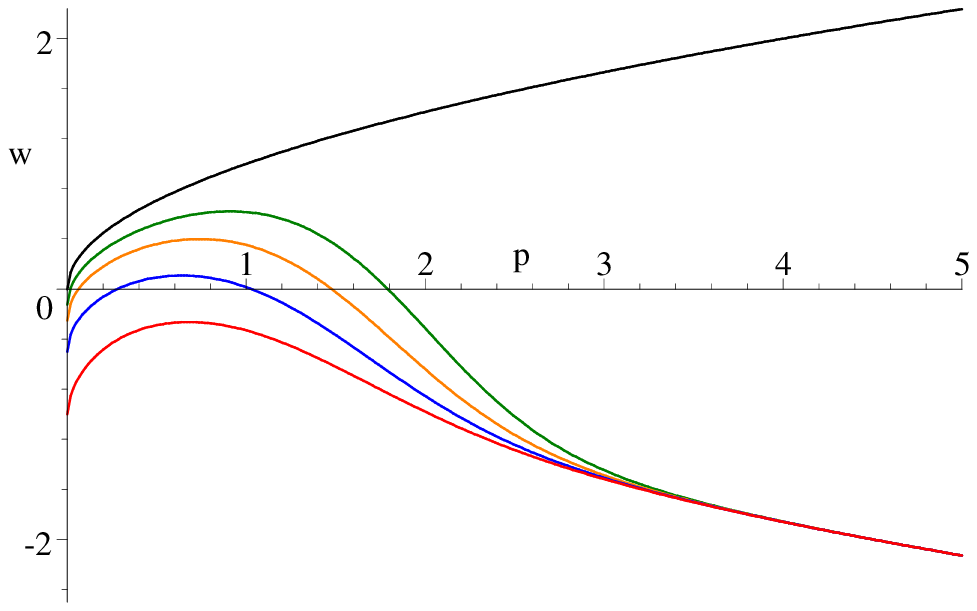}%
\\
$w_{\kappa}\left(  p\right)  $ for $\kappa=1,\ 1/2,\ 1/4,\ $\& $1/8$, in red,
blue, orange, \& green, respectively, along with $w_{0}\left(  p\right)
=\sqrt{p}$ in black.
\end{center}

The large $p$ behavior of $w_{\kappa}\left(  p\right)  $ for $\kappa\neq
0$\ may be seen analytically from the asymptotic behavior of (\ref{g(p)}),
which gives%
\begin{equation}
g\left(  p\right)  \underset{p\rightarrow\infty}{\sim}\frac{1}{2}%
\frac{e^{\frac{4}{3}p^{\frac{3}{2}}}}{\sqrt{p}}\left(  1+\frac{1}{2p^{\frac
{3}{2}}}+O\left(  \frac{1}{p^{3}}\right)  \right)  \ ,\ \ \ w_{\kappa}\left(
p\right)  \underset{p\rightarrow\infty}{\sim}\sqrt{p}\left(  -1+\frac
{1}{p^{\frac{3}{2}}}+O\left(  \frac{1}{p^{3}}\right)  \right)  \ .
\label{asymp}%
\end{equation}

As was the case for the undeformed model, there are some technical issues
associated with probability flow and self-adjointness of the Hamiltonians when
$\kappa\neq0$. \ We are content to leave these issues as exercises for the
interested reader.

\section{Discussion}

As emphasized by Shapere and Wilczek, \textquotedblleft many
worlds\textquotedblright\ systems with branched Hamiltonians are by no means
rare, in theory. \ Here, we have displayed some simple unified Lagrangian
prototype systems which, by virtue of non-convexity in their velocity
dependence, branch into double-valued (but still self-adjoint) Hamiltonians. \ 

We have outlined a gaussian model whose branches lie on a compact, closed
momentum manifold with coalescing cusps at finite $p$, as a preliminary step
in the search for a supersymmetric model with similar properties. \ We then
discussed a class of models with double-valued Hamiltons, one of which has the
canonical structure of a supersymmetric pair of Hamiltonians. \ We have
surveyed the spectral and boundary condition linkages involved across the
respective branches for this supersymmetric model, in a uniform framework, by
utilizing the eigenstate-linking supercharge ladder operators (but which are
\emph{not} Grassmann and which do \emph{not} commute with the two
Hamiltonians). \ 

These particular branched Hamiltonians --- although living in
\textquotedblleft two worlds\textquotedblright\ ---\ are nevertheless paired
by supercharges into a uniform Darboux isospectral system, in the very same
Hilbert space; and yet they are inexorably separated, in some analogy to
fermionic and bosonic sectors, as the respective dynamical intervals only
connect at $p=\infty$. \ In this respect, this particular supersymmetric
system differs from more typical constructions given by Shapere and Wilczek,
which exhibit similar operator branching structures but connect for finite
$p$.\newpage

\paragraph*{Acknowledgements:}

\noindent\textsl{{\small This work was supported in part by NSF Award
PHY-1214521; and in part, the submitted manuscript has been created by
UChicago Argonne, LLC, Operator of Argonne National Laboratory. Argonne, a
U.S. Department of Energy Office of Science laboratory, is operated under
Contract No. DE-AC02-06CH11357. The U.S. Government retains for itself, and
others acting on its behalf, a paid-up nonexclusive, irrevocable worldwide
license in said article to reproduce, prepare derivative works, distribute
copies to the public, and perform publicly and display publicly, by or on
behalf of the Government. \ TLC was also supported in part by a University of
Miami Cooper Fellowship.}}

\end{document}